\documentclass{article}

\usepackage{amsmath}
\usepackage{amssymb}
\usepackage{graphicx} 
\usepackage{geometry}
\usepackage{float} 
\usepackage{esvect}
\usepackage{setspace}
\usepackage[normalem]{ulem}
\usepackage{authblk}

\title{Single-Emitter Spectra from an Ensemble }

\author[1,$\perp$]{Jonah R. Horowitz}
\author[1,$\perp$]{Oliver J. Tye}
\author[1,2]{Oliver M. Nix}
\author[1]{Shaun Tan}
\author[3]{Hogeun Chang}
\author[3]{Ji Hyun Min}
\author[3]{Taehyung Kim}
\author[1,$\ast$]{Moungi G. Bawendi}

\affil[1]{\textit{Department of Chemistry, Massachusetts Institute of Technology, Cambridge, Massachusetts 02139, USA
}}
\affil[2]{\textit{Research Laboratory of Electronics, Massachusetts Institute of Technology, Cambridge, MA 02139, USA
}}
\affil[3]{\textit{Samsung Advanced Institute of Technology, Samsung Electronics, Suwon-si, Republic of Korea
}}

\date{} 

\begin{document}
\maketitle

\begin{center}
    \textsuperscript{$\perp$}\textit{These authors contributed equally to this work.}\\
    \textsuperscript{$\ast$}\textit{Corresponding author. Email: mgb@mit.edu}
\end{center}

\maketitle

\section*{Abstract}
The heterogeneity in nanoscale emitters hinders efforts to understand their basic photophysics and limits their use in practical applications. Existing methods have difficulty accurately characterizing single-emitter spectra and optical heterogeneity on a statistical scale. Here, we introduce SPICEE (SPectrally Imbalanced Correlations from Ensemble Emission), a spectrally filtered photon-correlation technique that recovers single-particle emission lineshapes from an ensemble sample. Analytical derivations, numerical modeling, and experiments on a solution ensemble of emitters validate the technique. We apply SPICEE to blue-emitting ZnSeTe semiconductor nanocrystals relevant to display applications and find that the low color purity in the ensemble spectrum is primarily caused by a small subpopulation of nanocrystals with a distinct emission mechanism. This work demonstrates that SPICEE is a powerful high-throughput tool for accurately characterizing the single-emitter properties of nanoscale systems.  

\section*{Main}
Nanoscale emitters are ubiquitous in modern science, finding diverse applications in displays, bioimaging, lasers, and quantum optics~\cite{jang_review_2023, lelek_single-molecule_2021, park_colloidal_2021,aharonovich_solid-state_2016}. Due to heterogeneities in size, morphology, and composition, optical properties can  vary greatly between individual emitters~\cite{cui_direct_2013,hayee_revealing_2020, moerner_dozen_2002, angell_unraveling_2024}. In turn, this optical heterogeneity broadens the ensemble emission spectrum, obscuring a material's fundamental single-emitter photophysics and limiting its applications. 

Existing methods struggle to quantify a sample's optical heterogeneity and single-emitter properties with both accuracy and high statistical significance. In conventional single-emitter spectroscopy, particles are spatially separated and probed one-by-one, making it difficult to collect spectra from a large number of individual emitters~\cite{moerner_dozen_2002}. Single-particle sample preparation and prolonged photoexcitation may also influence the spectral observables. As an alternative, ensemble-level techniques use nonlinear or correlative effects to infer single-emitter properties from the ensemble~\cite{cui_direct_2013, becker_long_2018} but in the process lose spectral information and average over variations between emitters. 

To address this problem, we present SPICEE (SPectrally Imbalanced Correlations from Ensemble Emission), a measurement that obtains population-resolved single-emitter spectra. These population-resolved spectra report on the variations in single-emitter properties within an ensemble with high statistical significance. We derive the relationship between measured intensity cross-correlations and single-emitter spectral properties. With computational modeling we show that the measurement is highly sensitive to the distribution of single-emitter spectra, rather than the globally averaged spectrum. Experimentally, we apply SPICEE to ensembles of semiconductor nanocrystals (NCs) that are of importance to heavy-metal-free quantum dot light-emitting diodes (QD-LEDs). In a sample of blue-emitting ZnSeTe NCs, SPICEE reveals the existence of subpopulations of NCs with distinct emissive mechanisms. Our results uncover a previously obscured regime of material properties that is critical to the understanding and application of nanoscale emitters.  

\begin{figure}[htbp]
	\centering
	\includegraphics[width=1\textwidth]{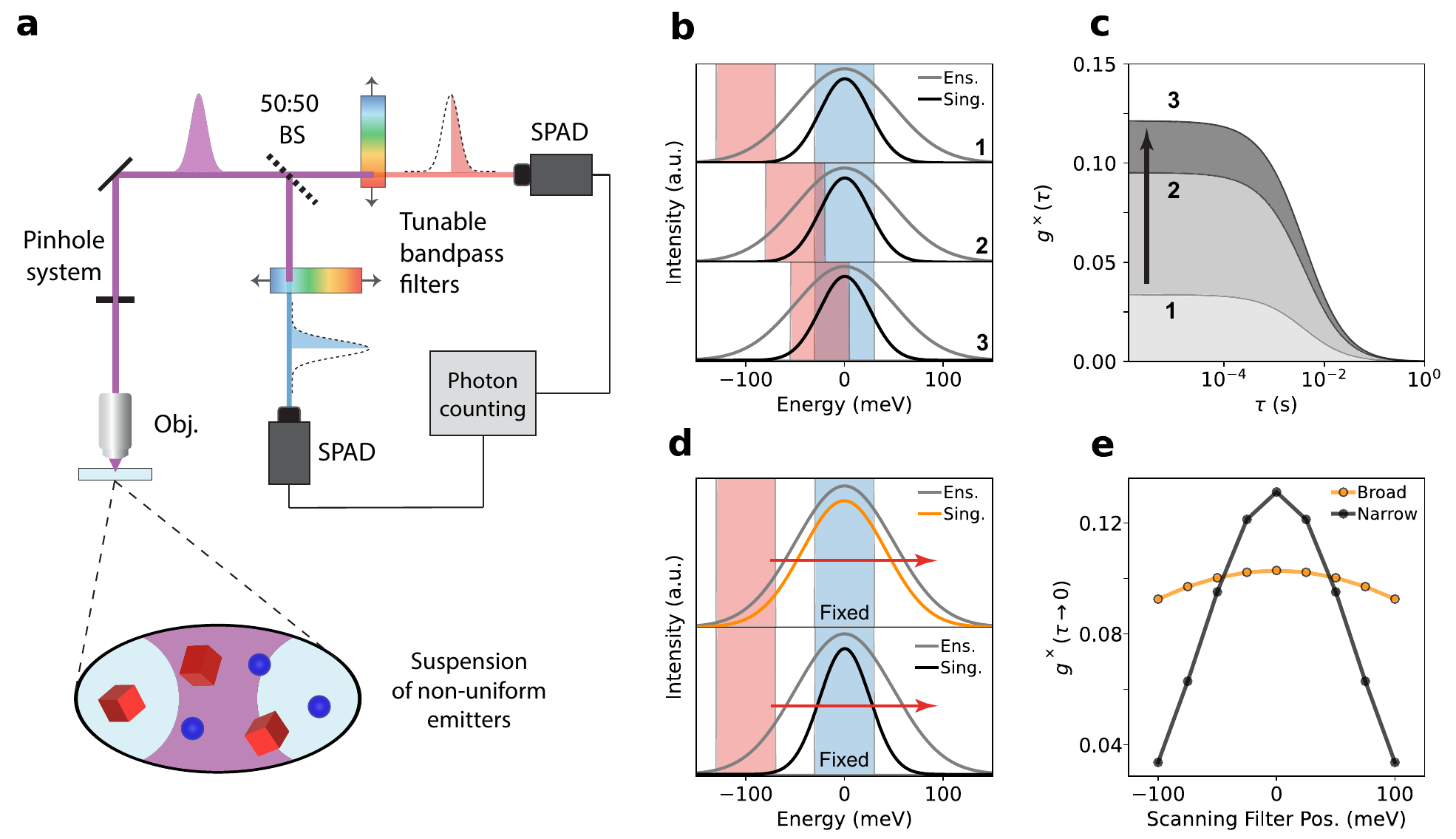}
    
	\caption{Illustration of the SPICEE technique.
         (\textbf{a}) Emission from a solution ensemble of emitters is collected by a confocal microscope and passed through a 50:50 beamsplitter (BS) and two independently tunable spectral bandpass filters before being collected at two single photon avalanche diodes (SPADs) Abbreviations: Obj., microscope objective. (\textbf{b}) Single-emitter (black) and ensemble (gray) spectra with select filter transmission profiles (red, blue). (\textbf{c}) Intensity cross-correlations $g^{\times}(\tau)$ of the detected emission (gray) for the corresponding filter positions in the previous panel. (\textbf{d}) Broad (orange) and narrow (black) single-emitter spectra with same ensemble (gray) are scanned over with the same filters. (\textbf{e}) The magnitude of the filtered $g^{\times}(\tau)$ at short $\tau$ is shown for filter positions from the scans in the previous panel. }

    \label{fig:fig1}
\end{figure}

\subsection*{Derivation and Modeling}

A SPICEE optical layout is shown in Figure 1a. Individual emitters in solution are probed with a low excitation flux while they diffuse through a laser focal spot on microsecond to millisecond timescales, as in fluorescence correlation spectroscopy (FCS)~\cite{webb_fluorescence_2001}. This excitation scheme efficiently samples from the entire ensemble while preventing photoinduced degradation. In an alternative experimental setup, the confocal laser spot could be scanned over a sparse solid-state sample~\cite{xiao_scanning_2005}, enabling SPICEE measurements on emitters in the solid state and at low temperatures. Dark-field techniques could also be used to study sample scattering~\cite{liu_scatter_2014}. In any of these implementations, the sample emission or scatter is then collected through an objective, spatially filtered by a pinhole aperture, split through a 50:50 beamsplitter, and directed towards two single photon avalanche diodes (SPADs). Before each detector, the light passes through a tunable spectral filter. These filters are tuned independently of one another and each transmit only a portion of the ensemble spectrum. 

Spectrally filtering the ensemble emission encodes single-emitter spectral information into the height of the background-subtracted cross-correlation $g^\times(\tau)$. As shown in the Supplementary Information (SI), the height of the SPICEE cross-correlation at short times is given by
\begin{equation}
    g^{\times}(\tau\rightarrow0)=\frac{1}{\langle N\rangle}
    \frac{\int_{-\infty}^{\infty} P(\mu)\Big[\int_{-\infty}^{\infty} s_{\mu} (\omega) f_A(\omega)\,d\omega \int_{-\infty}^{\infty} s_{\mu} (\omega)f_B(\omega) \,d\omega \Big] \, d\mu}{\Big[\int_{-\infty}^{\infty}  P(\mu)\int_{-\infty}^{\infty} s_{\mu} (\omega) f_A(\omega)\,d\omega \,d\mu \Big]\Big[\int_{-\infty}^{\infty} P(\mu)\int_{-\infty}^{\infty} s_{\mu} (\omega)f_B(\omega) \,d\omega \,d\mu\Big]  } .
\end{equation}
\noindent Here $\langle N\rangle$ is the average occupancy of the focal volume, and the remaining variables are spectral parameters. The numerator in the spectral factor gives the probability that two photons from a single emitter both pass through the two filters, while the denominator gives the probability that two separate photons from any two emitters are transmitted. The function $P(\mu)$ is the population distribution, corresponding to the probability that a single emitter has a peak energy of $\mu$. This quantitatively describes the ensemble-level broadening that occurs due to different emitters having different peak energies. The variable $s_\mu(\omega)$ is the average single-emitter spectrum (written as a function of energy $\omega$) for an emitter with a peak energy of $\mu$. Therefore, the SPICEE response is sensitive to the evolution in the single-emitter spectrum as a function of peak energy. Finally, $f_{A/B}(\omega)$ are the transmission profiles of the two spectral filters. 

The evolution of the SPICEE $g^{\times}(\tau)$ as a function of time delay contains additional information. Owing to particle diffusion, the $g^{\times}(\tau)$ decays with the characteristic time constant $\tau_\mathrm{D}$, corresponding to the average dwell time in the focal volume, similar to FCS~\cite{yu_comprehensive_2021}. If the emitters undergo spectral diffusion on timescales $0<\tau\le\tau_D$, the shape of  the correlation function is altered, because over time the individual emitter may enter or leave the spectral detection window (more details in the SI). Rapid spectral diffusion appears to be negligible compared to the linewidth in solution-bound colloidal semiconductor NCs~\cite{marshall_extracting_2010, utzat_probing_2017}. However, SPICEE may be used to study the spectral dynamics in other nanoscale systems, such as fluorescent proteins or single-photon emitters.

We use computational modeling to illustrate the mathematical relationship between $g^{\times}(\tau)$ and the single-emitter spectra. In Figure 1b, we model a simple system in which all emitters exhibit the same spectral lineshape but the distribution in peak energies results in an ensemble spectrum (gray) that is broader than any single-emitter spectrum (example in black). We assume that the system has no spectral dynamics. We then introduce two bandpass filters with boxcar-function transmission profiles (red, blue). As the filters are brought closer together, the spectral overlap between the filters and a given single emitter increases (Fig. 1b). As a result, the height of $g^\times(\tau)$ rises (Fig. 1c). The response in the $g^\times(\tau)$ depends on the linewidths of the single-emitter spectra. In Figure 1d, we show two model systems with the same ensemble spectrum (gray) but either broad (orange) or narrow (black) single-emitter spectra. We sweep one of the bandpass filters (red) while holding the other (blue) constant. The two systems yield different responses in the heights of the $g^\times(\tau)$ (Fig. 1e), demonstrating that the SPICEE observable is sensitive to the spectral properties of single emitters. It is worth noting that the evolution of $g^\times(\tau)$ as a function of filter position does not directly correspond to the single-emitter lineshape, but also depends on the population distribution and, as we will show, the variation in the single-emitter lineshape. SPICEE can extract all of these parameters simultaneously, distinguishing it from previous techniques.

\begin{figure}[htbp]
	\centering
	\includegraphics[width=1\textwidth]{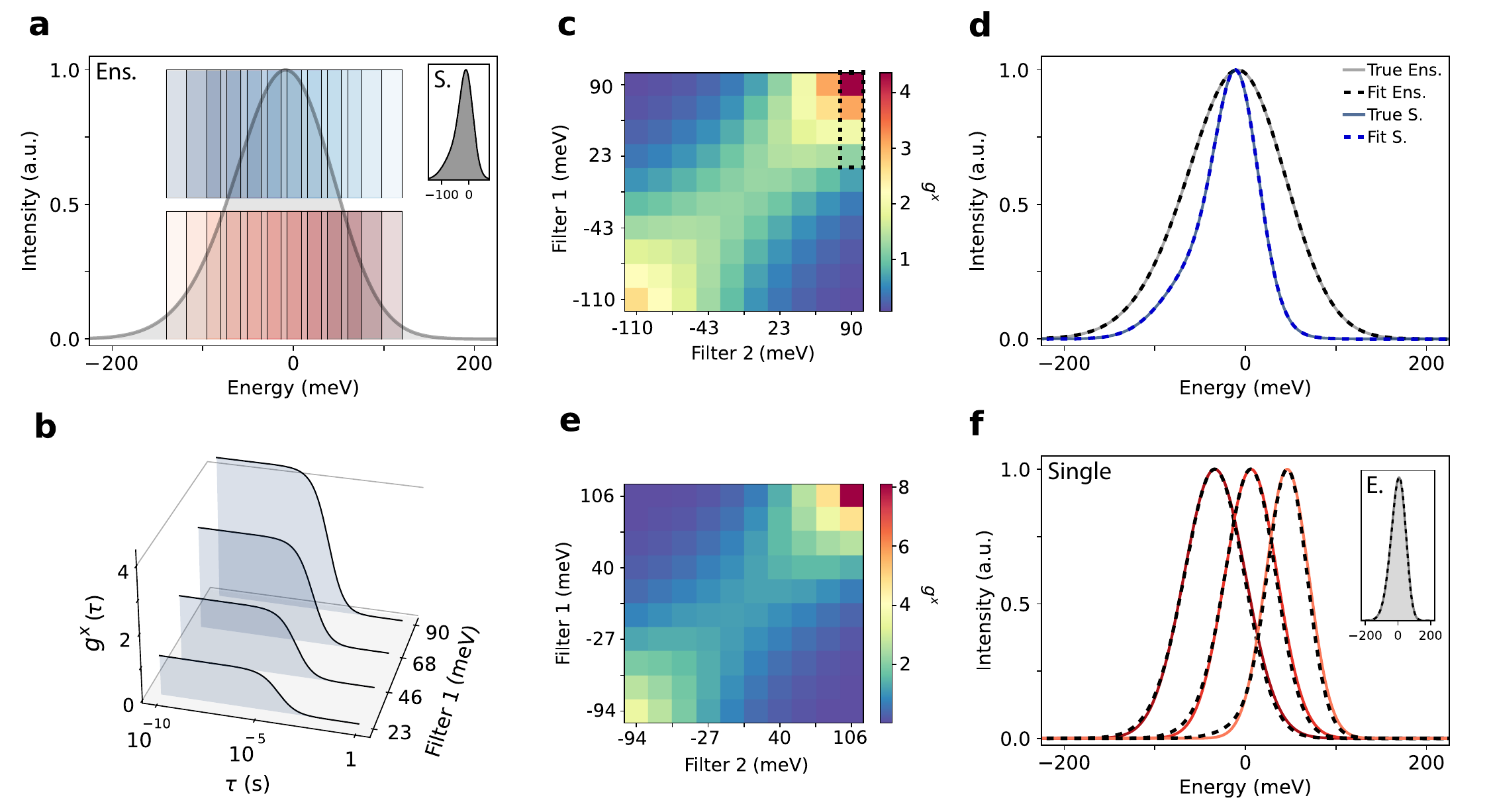}
    
	\caption{SPICEE resolves asymmetric lineshapes and lineshape evolution in model systems. 
         (\textbf{a}) Ensemble spectrum (gray) overlaid with the two sets of 10 spectral filters (red, blue) with offset for clarity. The single-emitter spectrum (inset, gray) shows a clear asymmetry. (\textbf{b}) $g^\times(\tau)$ for select filter positions where $\langle N\rangle =1$. (\textbf{c}) $g^\times(\tau\to0)$ for each set of filter positions, creating a 10 x 10 grid. (\textbf{d}) Fitted SPICEE grid recovers the single emitter and ensemble spectrum with high fidelity. (\textbf{e}) SPICEE grid for a system where the spectral linewidth evolves as a function of peak position. (\textbf{f}) Fitting the SPICEE grid from the previous panel recovers the single-emitter spectral evolution, as well as the ensemble spectrum (inset).
    }
	\label{fig:fig2}
\end{figure}

To quantitatively determine the single-emitter spectra and population distribution, we fit the SPICEE data, using Equation 1 and analytical expressions for each function (details in the SI). To test the ability of this procedure to extract the spectral parameters, we model systems with more complicated spectral properties, which would be inaccessible to other methods that sample from the ensemble. We start with an inhomogeneously broadened system where each single-emitter spectrum exhibits the same pronounced asymmetry towards lower energies (Fig. 2a, inset). We model a SPICEE experiment for this system, stepping two sets of filters (both 60 meV-wide boxcar functions) over a series of ten positions (Fig. 2a, blue, red) covering the ensemble spectrum (gray). For every combination of filter positions, we apply Equation 1 to obtain the cross-correlation height (taking $\langle N\rangle=1$). Complete $\tau$-dependent cross-correlations for select positions are shown in Figure 2b. Taking the cross-correlation heights from all filter pairs produces a 10 by 10 grid (Fig. 2c), where a row or column in the grid would correspond to the one-dimensional filter sweeps shown in Figure 1e. As in Figure 1e, each row/column peaks where the filters are maximally overlapped, creating a clear diagonal pattern in the grid. The high- and low-energy extrema along the diagonal give the largest values in the dataset, because those positions correspond to the tails of the ensemble spectrum, where the concentration of emitters is the lowest and the $g^\times(\tau)$ has the greatest contrast. We fit the data using Equation 1 to extract the universal single-emitter lineshape and population distribution. Figure 2d shows excellent agreement between the inputted and fitted data for both the single-emitter and ensemble spectra. These results demonstrate that fitting to SPICEE data can recover detailed single-emitter spectral information, including asymmetries in the lineshape. 

In certain nanoscale systems, such as NCs with a mixture of excitonic and trap emission~\cite{lee_coherent_2023}, the single-emitter spectrum may evolve as a function of peak energy. SPICEE can resolve these emitter-to-emitter variations, distinguishing it from other ensemble-level techniques~\cite{cui_direct_2013, becker_long_2018}. To demonstrate this novel ability, we model a system in which the single-emitter linewidth transitions from narrow (FWHM=40 meV) at high energies to broad (FWHM=100 meV) at low energies. The modeled SPICEE data are shown in Figure 2e and are fit using the procedure described in the SI. We observe strong agreement between the fit and inputted 
spectra for the single emitters as well as the ensemble (Fig. 2f). This result confirms that fitting to the SPICEE data can quantitatively recover an evolving single-emitter lineshape.

\subsection*{Experimental Results}

To test SPICEE experimentally, we constructed the optical setup given by the schematic in Figure 1a and synthesized a batch of InP/ZnSe/ZnS NCs according to Won \textit{et al.}~\cite{won_highly_2019}. These NCs produce an ensemble photoluminescence (PL) spectrum with a maximum at 624 nm and a FWHM of 35 nm (110 meV) (Fig. 3a). For this sample, we measure a photoluminescence quantum yield (PLQY) of 80\%. The NCs have a core-shell-shell architecture, with average dimensions of 3.3 nm for the InP core diameter, 2.7 nm for the inner ZnSe shell thickness, and 0.15 nm for the outer ZnS shell thickness~\cite{won_highly_2019}. The sample's spectral purity and heavy metal-free composition make it a leading candidate for electrically driven QD-LEDs. 

\begin{figure}[htbp] 
	\centering
	\includegraphics[width=1\textwidth]{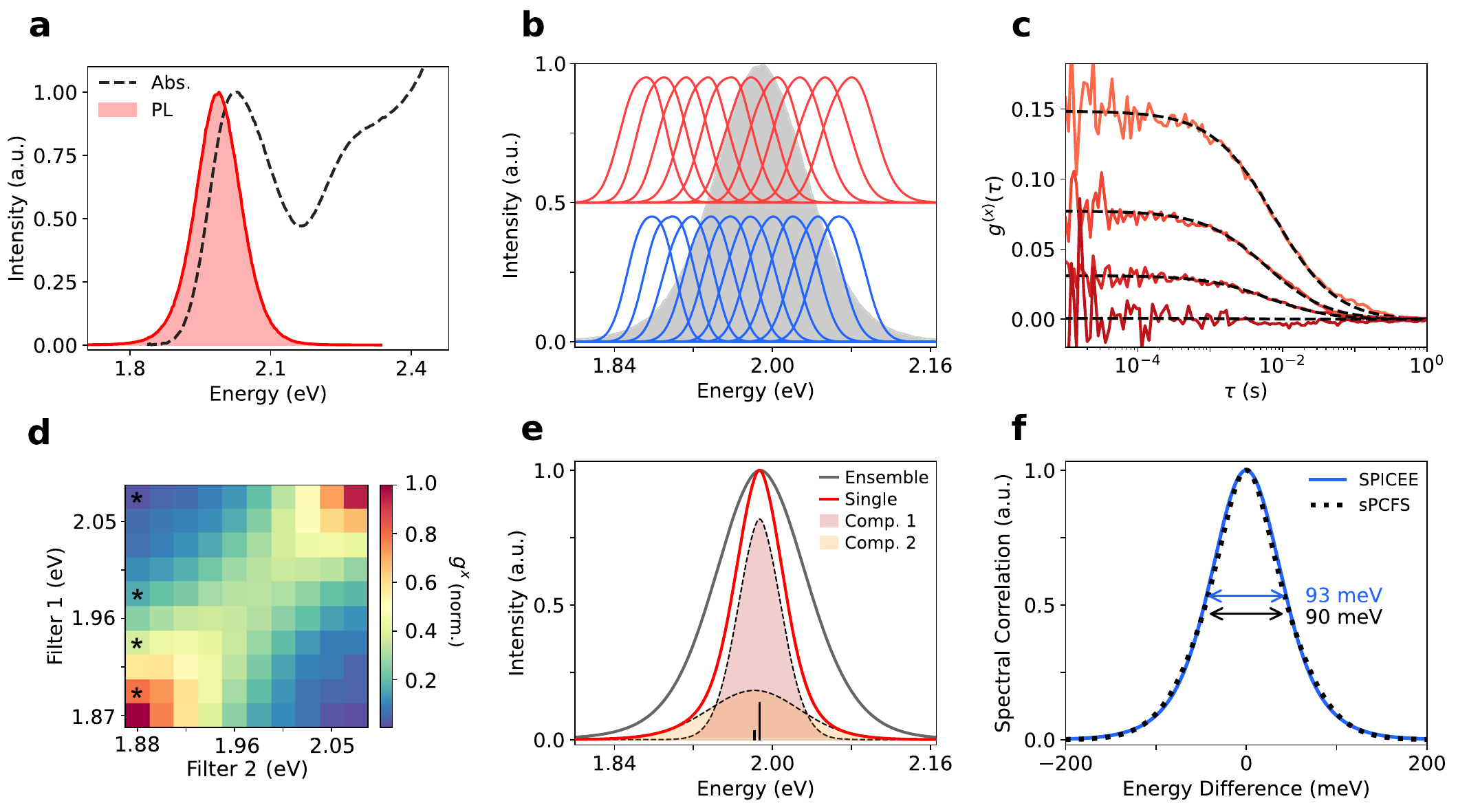}
    
	\caption{SPICEE experiment on a solution ensemble of InP/ZnSe/ZnS NCs.
          (\textbf{a}) Ensemble absorption and PL spectra for InP/ZnSe/ZnS.  (\textbf{b}) Normalized spectral filter transmission profiles (red, blue, 10 each, vertical offset for clarity) relative to the ensemble spectrum (gray, shaded). (\textbf{c}) $g^{\times}(\tau)$ as one filter is moved from closely overlapped with the other (light red) to far apart (dark red) with fits (black). (\textbf{d}) The normalized SPICEE grid for all combinations of filters with the corresponding $g^\times$ positions in the previous panel marked with asterisks (*). (\textbf{e}) Fitted single-NC spectrum (red) and its components (dashed) with centers and relative weights of the spectral components as black bars. The ensemble spectrum is shown in gray. (\textbf{f}) Comparison to sPCFS spectral correlation shows excellent agreement between the two measurements.
    }
	\label{fig:fig3}
\end{figure}

Next, we used SPICEE to measure the single-emitter spectral properties of a solution ensemble of the InP NCs. Figure 3b shows the transmission profiles of the filters for this experiment (red, blue), overlaid on the ensemble spectrum (gray). Several measured $g^\times(\tau)$ from the experiment are shown in Figure 3c (corresponding to the boxes marked with asterisks (*) in Fig. 3d). As expected, the height of the $g^\times(\tau)$ dramatically varies with the filter position. We fit each $g^\times(\tau)$ to extract the value at short $\tau$. We repeat this process for every combination of filters, creating a 10 by 10 grid, which we normalize to the maximum value (Fig. 3d). We correct for any long-term drift in the total NC concentration using the autocorrelations (SI). Given the sample's high PLQY and largely symmetric ensemble spectrum, we fit the SPICEE results to a static single-NC spectrum, assuming a double Gaussian lineshape. We fit the population distribution to a single Gaussian, having found no improvement in the fit using a double Gaussian (SI). After obtaining an initial fit, we apply a Monte Carlo (MC) error analysis~\cite{press_numerical_1993}, to rigorously estimate the fit parameter errors (more details in the SI). We average the MC fits to obtain the most-likely single-NC spectrum (Fig. 3d). It has a FWHM of 58 meV, which is consistent with single-particle studies~\cite{berkinsky_narrow_2023}. The spectrum also has a slight but statistically significant red tail (average offset = -8.6 meV), likely due to InP and/or ZnSe optical phonons~\cite{berkinsky_narrow_2023}. We determine that the most-likely population distribution has a FWHM of 93 meV, implying emitter heterogeneity is a significant contributor to the ensemble linewidth. Although techniques such as transmission electron microscopy (TEM) may be used to infer distributions in structural parameters across particles, these data can be difficult to connect to the spectral observables of interest. SPICEE therefore provides a more direct assessment of the optical heterogeneity in synthetic nanomaterials. 

To test the accuracy of SPICEE, we measured the same batch of InP NCs using solution photon-correlation Fourier spectroscopy (sPCFS) \cite{brokmann_revealing_2009, marshall_extracting_2010}, an ensemble-level interferometric method which provides globally averaged single-emitter properties. For consistency, we used the same solution sample preparation and laser excitation conditions. Unlike SPICEE, sPCFS does not directly report on the single-emitter spectrum. Instead, it yields the symmetrized autocorrelation of the single-emitter spectrum, or spectral correlation. To obtain the experimental spectral correlation for the InP NCs, we fit the sPCFS data assuming a double-Gaussian single-NC spectrum. We again applied a MC process to estimate the error in the sPCFS spectral parameters (details in the SI). To properly compare the results of the two measurements, we generate a spectral correlation from the most-likely SPICEE single-NC spectrum. Figure 3f shows that the most-likely SPICEE and sPCFS spectral correlations are in excellent agreement. The linewidths of the spectral correlations are very similar: the SPICEE spectral correlation has a FWHM of 93 meV (95\% CI: 77 - 98), while the sPCFS spectral correlation has a FWHM of 90 meV (95\% CI: 86 - 94). This agreement convincingly validates the SPICEE method.

Finally, we apply SPICEE to a solution ensemble of ZnSeTe-based NCs. ZnSeTe NCs have emerged in recent years as a leading heavy-metal-free material for blue electrically-driven QD-LEDs~\cite{kim_efficient_2020,  wu_homogeneous_2025, jang_synthesis_2019, han_more_2020}. The primary obstacles to the commercial implementation of ZnSeTe-based QD-LEDs are poor device stabilities, as well as low spectral purities due to the introduction of Te \cite{tan_challenges_nodate}. A deeper understanding of the single-NC photophysics of these materials may enable improved device performances. Suggested mechanisms for the low spectral purity include: non-uniform shifting of the exciton bandgap due to heterogeneous doping levels~\cite{jang_synthesis_2019, cai_emission_2024}; the increased density of lattice defects~\cite{kim_efficient_2020, wu_homogeneous_2025, jang_synthesis_2019}; and the formation of hole-localizing mid-gap states from clusters of Te atoms~\cite{kim_efficient_2020, wu_homogeneous_2025, imran_molecular-additive-assisted_2023, cai_emission_2024}. In this work, we synthesized a batch of core-shell-shell ZnSeTe/ZnSe/ZnS NCs by modifying a procedure first described in Kim \textit{et al.}~\cite{kim_efficient_2020}. The ZnSeTe core was synthesized with a 2.5\% Te/Se molar ratio, with an average diameter of 3.1 nm. The average thicknesses of the ZnSe inner shell and ZnS outer shell are 2.6 nm and 1.2 nm, respectively. Optically, we measure a PLQY of 80\%. The ensemble PL spectrum (Fig. 4a) shows a peak wavelength of 443 nm and a narrow FWHM of 16 nm (100 meV). But, as in other ZnSeTe NCs, the ensemble PL shows a pronounced low-energy tail that approaches green wavelengths, diminishing the sample's spectral purity. 

\begin{figure}[htbp] 
	\centering
	\includegraphics[width=1\textwidth]{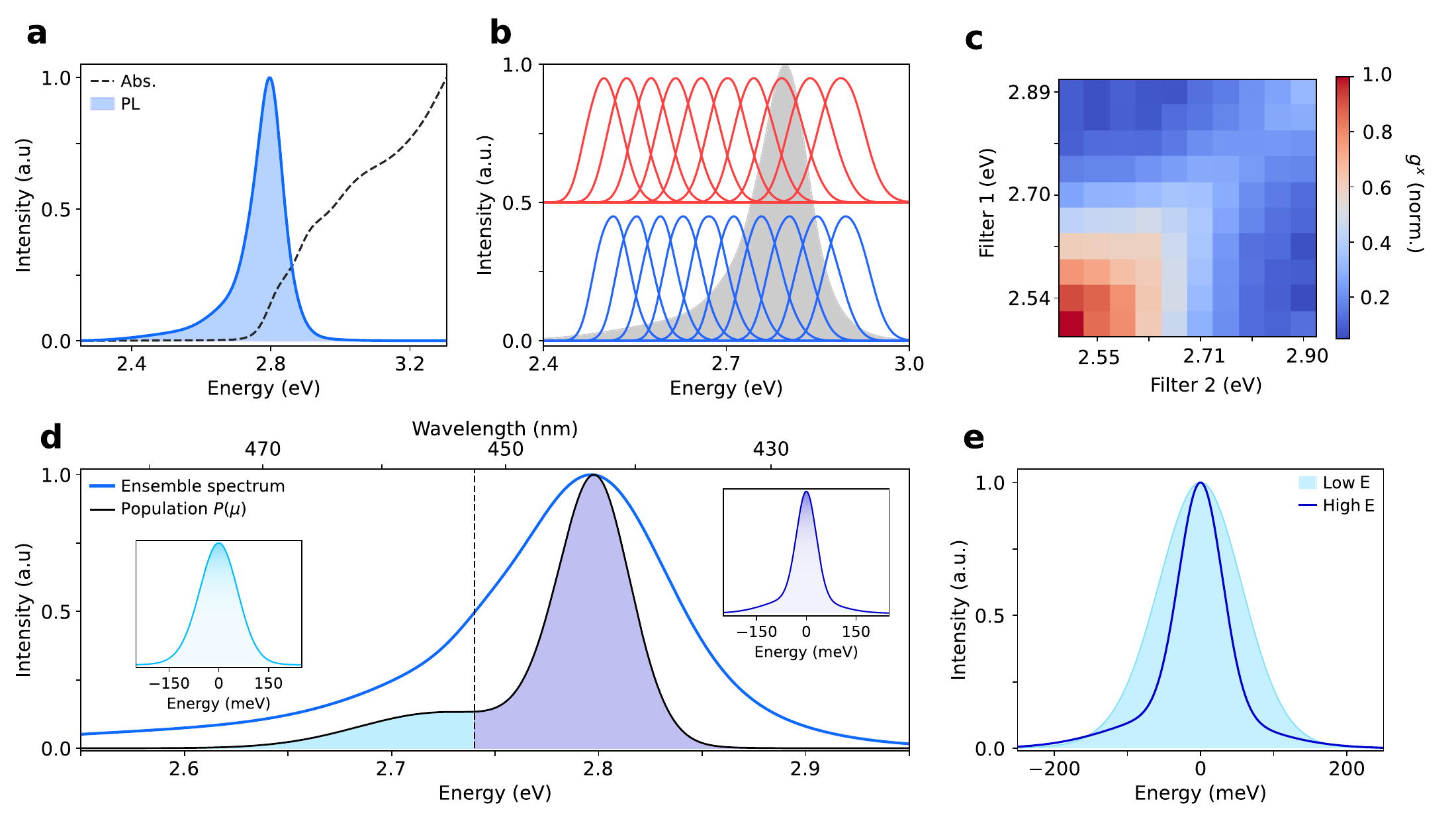} 
    
	\caption{SPICEE applied to a solution ensemble of ZnSeTe/ZnSe/ZnS NCs. 
        (\textbf{a}) Ensemble absorption (gray, dotted) and PL (blue, shaded). (\textbf{b}) Normalized filter transmission profiles relative to the ensemble spectrum. (\textbf{c}) Normalized SPICEE grid. (\textbf{d}) Spectral decomposition of the ZnSeTe sample. The ensemble spectrum (blue) is composed of two subpopulations of NCs (relative weights in black) where the dominant population has a narrow 80 meV linewidth and the smaller population has a broader 140 meV linewidth. (\textbf{e}) Overlay of the broad, low-energy single-NC spectrum and the narrow, high-energy single-NC spectrum.
    }
	\label{fig:fig4} 
\end{figure}

We performed SPICEE on a solution ensemble of these ZnSeTe NCs. The experimental filters for this measurement (red, blue), overlaid on the ensemble spectrum (gray), are shown in Figure 4b. There are ten filters for each path, for a total of one hundred combinations of filter functions. The measured $g^\times(\tau\rightarrow0)$ values are plotted in a grid in Figure 4c. The $g^\times(\tau\rightarrow0)$ values are greater on the low-energy corner (bottom left) than at the high-energy corner (upper right), indicating that the number of emitters near the red tail of the spectrum is lower than near the peak. This result suggests that there may be multiple populations of spectrally distinct NCs within the ensemble. To capture the potential evolution of the single-NC spectra without a highly parameterized model, we split the energy axis into two intervals and separately fit the average single-NC spectrum on either side of the cutoff. We identified the cutoff, $E_c = $ 2.74 eV (452 nm), by finding the energy within a reasonable range that gave the lowest fit error. Within this range, the cutoff energy did not significantly impact the fitted single-NC spectra or population distribution.

Figure 4d summarizes the results of the fit, showing the population distribution overlaid with the average single-NC spectra corresponding to the high- and low-energy regions. We see that the population distribution $P(\mu)$ is composed of 1) a narrow, high-amplitude component with a peak at 2.80 eV (443 nm) and 2) a broad and low shoulder on the low-energy side extending about -150 meV (p95) beneath the peak of the distribution. We find that the NCs that correspond to the dominant, narrow distribution (with peak energies $>$ 2.74 eV) have relatively narrow single-particle spectra: the average spectrum for the high-energy interval has a most-likely FWHM of about 80 meV. In contrast, the NCs making up the shoulder in the population distribution (with peak energies $\leq$ 2.74 eV) have broad single-particle spectra: the average spectrum for the low-energy interval has a most-likely FWHM of about 140 meV. Figure 4e overlays the pair of average single-NC spectra, highlighting the striking difference in linewidths. From these results we conclude that the pronounced asymmetry in the ensemble PL spectrum is primarily caused by the redshifted population. These NCs have linewidths that are on average about 1.8 times broader than those of the unshifted population, further contributing to spectral broadening. Recalling that localized trap emission is generally red-shifted and broadened relative to excitonic emission~\cite{empedocles_photoluminescence_1999, mooney_challenge_2013, hinterding_single_2021}, our results are consistent with a primary population of NCs that emit mainly from the exciton and a secondary population of NCs that emit mainly from trap states. The broad pedestal in the average high-energy single-NC spectrum (Fig 4e, dark blue) may reflect a fraction of the emission coming from relatively shallow traps. Based on our fit, we make the estimate that roughly 15\% of the NCs emit from the deeper trap states with peak energies redshifted by $\geq$ 60 meV relative to the population maximum.

Previous studies have suggested that ZnSeTe NCs suffer from localized emission, likely due to clusters of Te atoms~\cite{cai_emission_2024, wu_homogeneous_2025, imran_molecular-additive-assisted_2023, lee_crystallographic_2024, kim_efficient_2020}. However, our results mark some of the clearest single-NC observations to date of distinct populations of excitonic and localized emitters in ZnSeTe NCs. Previously, Wu \textit{et al.} had collected PL spectra of single ZnSeTe NCs, generally observing that redshifted NCs had broader spectra~\cite{wu_homogeneous_2025}. However, these measurements were performed on unshelled cores and showed extremely broad FWHMs (generally above 200 meV), making this trend difficult to interpret. Interestingly, Imran \textit{et al.} observed little variation in the PL spectra of individual ZnSeTe/ZnSe/ZnS NCs~\cite{imran_molecular-additive-assisted_2023}. The discrepancy between the results of Imran \textit{et al.} and those presented in this work may be attributable to synthetic differences. Imran \textit{et al.} also only report a relatively small number of single-NC spectra per sample, whereas SPICEE collects signal from a statistical number of emitters. Our results indicate a transition from a primarily excitonic sub-population to a primarily trapped one, consistent with hot charges rapidly cooling to the lowest-lying energetic state. Our findings on the single-NC emission pathways of this key nanomaterial will inform ongoing efforts to commercialize QD-LEDs. 

In the future, SPICEE may be applied to a wide range of nanoscale emitters, including molecules, fluorescent proteins, and color centers in crystals, as well as nanoscale scattering centers such as noble-metal nanoparticles. Using scanning excitation, SPICEE can also be performed on materials in the solid state at cryogenic temperatures. Given the measurement's minimal user input, it could be effectively integrated into automated high-throughput workflows. The generality of SPICEE, combined with its efficiency and high statistical rigor, position it as a promising new technique for use across photonics research.

\section*{Methods}

\subsection*{InP/ZnSe/ZnS Synthesis}

The InP/ZnSe/ZnS NCs studied in this work were synthesized according to the procedure for ``QD-2R'' in Won \textit{et al.}~\cite{won_highly_2019}

\subsection*{ZnSeTe/ZnSe/ZnS Synthesis}
The ZnTeSe/ZnSe/ZnS NCs with 2.5\% Te investigated in this study were synthesized using a procedure modified from that described in Kim \textit{et al}.~\cite{kim_efficient_2020}. For this work, the ZnTeSe core was synthesized with a different amount of 0.1 M Te-TOP: we used 0.1125 mmol instead of 0.3 mmol. The ZnSe/ZnS double shell was grown on the core using the same method described in the previous report. The resulting nanocrystals were isolated using ethanol as an antisolvent and finally redispersed in octane for later use.

\subsection*{Ensemble Absorbance and PL Characterization}
Samples were prepared by diluting the NCs by a factor of 500 in toluene. Absorption spectra were measured with an Agilent Technologies Cary 5000 UV–vis–NIR spectrophotometer. PL spectra were measured using a HORIBA Jobin Yvon Fluoromax 3 spectrofluorometer.

\subsection*{Photoluminescence Quantum Yield (PLQY) Measurements}
The NC PLQYs were measured on a home-built optical setup. The samples were diluted roughly one-hundredfold in the same solvents used for the solution photon-correlation measurements (described below) and mounted in a Labsphere integrating sphere. The excitation source was a 405 nm CW laser (Thorlabs \#CPS405) focused with a lens (Thorlabs \#LA4545) onto the solution sample. The resulting emission, along with scattered laser light, was collected through the bottom of the integrating sphere and collimated using an off-axis parabolic mirror followed by a plano-convex lens (Thorlabs \#MPD00M9-F01 and \#LA4545, respectively). Finally, the light was focused (Thorlabs \#LA4855) into a spectrometer (Princeton Instruments, SP2300i) paired with a CCD camera (Princeton Instruments, PIXIS 100 BRexcelon). We corrected for the wavelength-dependent response of the setup by also measuring an intensity-calibration lamp with a known spectrum (Ocean Optics, HL-3 plus -CAL-INT). The PLQYs were calculated using the method reported in de Mello \textit{et al.}~\cite{de_mello_improved_1997}. For the ZnSeTe sample, which showed signs of reabsorption, we corrected the measured PLQY using the approach described in Ahn \textit{et al.}~\cite{ahn_self-absorption_2007}.

\subsection*{NC solution sample preparation for correlation measurements}

All sample preparation for the sPCFS and SPICEE experiments was performed in a nitrogen glovebox with a measured oxygen concentration of \textless 0.1 ppm. The InP/ZnSe/ZnS NCs were diluted between five- and twenty-thousandfold in a solution of anhydrous toluene (Sigma-Aldrich \#244511), 1\% poly(methyl methacrylate) (PMMA, average Mw $\approx$ 120,000, Aldrich \#182230) by weight, and 0.1\% trioctylphosphine (TOP, Sigma-Aldrich \#718165) by volume. The stock solution had been previously dried over 4 \AA\ molecular sieves (Sigma-Aldrich \#334308). Similarly, for the ZnSeTe measurements, the NC stock solution was diluted fifteen-hundredfold in a solution of toluene, 1\% PMMA by weight, and 2\% TOP by volume, dried over molecular sieves. The manufacturers and part numbers for these materials are the same as those listed for the InP samples. The dilute solution of InP or ZnSeTe NCs was then drawn into a borosilicate glass capillary (VitroCom \#5012) capped with plastic putty (Leica Critoseal \#39215003).

\subsubsection*{sPCFS Optical Setup and Experiment}

The sPCFS experiments described in the main text were performed using a home-built confocal microscope, with a 405 nm CW diode laser (Melles Griot 56RCS/S2780) as the excitation source. First, the excitation was passed through a water immersion objective (Nikon Plan Apo VC 60XA/1.20 WI) and focused down on the NC solution sample. The capillary containing the sample was mounted perpendicularly to the optical axis. The resulting epifluorescence from the sample was then passed through a 405 nm dichroic to filter out the excitation light, leaving only the NC emission. Next, the emission was focused to a point using an achromatic lens (Thorlabs \#AC254-080-A-ML), spatially filtered through a 100 $\mu$m pinhole (Thorlabs \#P100HK), and re-collimated (same as focusing lens). Following the pinhole system, the NC emission was directed through a 750 nm shortpass filter to reduce the amount of background light. Next, the emission was sent through a Michelson interferometer equipped with a 50:50 cube beamsplitter (Thorlabs \#BS031). The position of the mobile arm of the interferometer was adjusted with a linear nanopositioning stage (Aerotech). The NC emission interference was then collected on a pair of  single-photon avalanche diodes (custom Excelitas AQRH-14 SPADs). The resulting photon streams were recorded using a Swabian TimeTagger 20 with 34 ps jitter. For the sPCFS experiment on InP NCs described in the main text, we set the laser power at $\approx$120 nW. We scanned the mobile arm of the Michelson interferometer over 61 stage positions centered approximately on the point of maximum interference. At the true point of maximum interference, or white fringe, the path length difference is equal to zero. The greatest path length difference measured was 37.5 $\mu$m. At every stage position, the linear stage dithered according to a triangular waveform with an amplitude of 400 nm and frequency of 0.067 Hz. We collected signal for four minutes at each stage position. We discuss the data analysis for this experiment at length in the SI.

\subsection*{SPICEE Optical Setup and Experiment}

The SPICEE experiments discussed in the main text were performed on the same home-built confocal microscope as the sPCFS measurements, with certain key changes. Again, all experiments were performed with the same 405 nm CW laser for the excitation source. Moreover, the entire excitation path, sample mounting, and emission path through the pinhole system were the same as described in the previous section. However, the experimental setups differed following the pinhole system. In the SPICEE experiments, after the pinhole the NC emission was split into two paths using a 50:50 plate beamsplitter (Thorlabs \#BSW10). Each path was then directed through a separate linear tunable bandpass filter (Edmund Optics Linear Variable Bandpass VIS \#88-365). Both filters were mounted on linear stages with 60 mm travel (Thorlabs Elliptec ELL20K), allowing for automated control of the filter profile. Finally, the emission was collected on a pair of single-photon avalanche diodes (customized Excelitas AQRH-14 SPADs). The detector apertures were covered with 750 nm shortpass filters to limit contributions from background light. As before, the photon stream was processed by a Swabian TimeTagger 20 module. The laser power was set to $\approx$120 nW for the measurement on the InP NCs and $\approx$700 nW for the ZnSeTe NCs. The total experimental times were around 5 and 5.5 hours for the InP and ZnSeTe samples, respectively. After initially loading the samples, the experiments ran without further user input.

\section*{Acknowledgements}
J.R.H., O.J.T., and S.T. acknowledge support from the Samsung Advanced Institute of Technology (SAIT). O.M.N. was supported by the US Department of Energy, Office of Science, Office of Basic Energy Sciences, Materials Chemistry Program under award number DE-FG02-07ER46454. We gratefully acknowledge Dr. Gang Liu for building electronics used in the SPICEE optical setup.

\section*{Author Contributions}
J.R.H. and O.J.T. contributed equally to this work. J.R.H. co-led the study, developed experimental control software, performed InP measurements, and contributed to analysis and interpretation of all data. O.J.T. co-led the study, developed SPICEE simulation/analysis software, performed ZnSeTe measurements, and contributed to analysis and interpretation of all data. O.M.N. assisted with quantum yield measurements. S.T. assisted with data analysis. J.M. synthesized the InP NCs and provided TEM images. H.C. synthesized the ZnSeTe NCs and provided TEM images. T.K. and  M.G.B. supervised the study.

\section*{Conflict of Interest}
MIT has filed a provisional patent application for the SPICEE technique that names J.R.H., O.J.T., and M.G.B. as inventors.

\bibliographystyle{unsrt}
\bibliography{references}

@article{liu_scatter_2014,
    title = {Tempo-Spatially Resolved Scattering Correlation Spectroscopy under Dark-Field Illumination and Its Application to Investigate Dynamic Behaviors of Gold Nanoparticles in Live Cells},
    volume = {136},
    number = {7},
    journal = {Journal of the American Chemical Society},
    author = {Liu, Heng and Dong, Chaoqing and Ren, Jicun},
    year = {2014},
    pages = {2775--2785},
}

@article{wu_homogeneous_2025,
    title = {Homogeneous {ZnSeTeS} quantum dots for efficient and stable pure-blue {LEDs}},
    volume = {639},
    copyright = {2025 The Author(s), under exclusive licence to Springer Nature Limited},
    number = {8055},
    journal = {Nature},
    author = {Wu, Qianqian and Cao, Fan and Yu, Wenke and Wang, Sheng and Hou, Wenjun and Lu, Zizhe and Cao, Weiran and Zhang, Jiaqi and Zhang, Xiaoyu and Yang, Yingguo and Jia, Guohua and Zhang, Jianhua and Yang, Xuyong},
    year = {2025},
    pages = {633--638},
}

@article{kim_efficient_2020,
    title = {Efficient and stable blue quantum dot light-emitting diode},
    volume = {586},
    copyright = {2020 The Author(s), under exclusive licence to Springer Nature Limited},
    number = {7829},
    journal = {Nature},
    author = {Kim, Taehyung and Kim, Kwang-Hee and Kim, Sungwoo and Choi, Seon-Myeong and Jang, Hyosook and Seo, Hong-Kyu and Lee, Heejae and Chung, Dae-Young and Jang, Eunjoo},
    year = {2020},
    pages = {385--389},
}

@article{won_highly_2019,
    title = {Highly efficient and stable {InP}/{ZnSe}/{ZnS} quantum dot light-emitting diodes},
    volume = {575},
    copyright = {2019 The Author(s), under exclusive licence to Springer Nature Limited},
    number = {7784},
    journal = {Nature},
    author = {Won, Yu-Ho and Cho, Oul and Kim, Taehyung and Chung, Dae-Young and Kim, Taehee and Chung, Heejae and Jang, Hyosook and Lee, Junho and Kim, Dongho and Jang, Eunjoo},
    year = {2019},
    pages = {634--638},
}

@article{lee_coherent_2023,
    title = {Coherent heteroepitaxial growth of {I}-{III}-{VI2} {Ag}({In},{Ga}){S2} colloidal nanocrystals with near-unity quantum yield for use in luminescent solar concentrators},
    volume = {14},
    copyright = {2023 The Author(s)},
    number = {1},
    journal = {Nature Communications},
    author = {Lee, Hak June and Im, Seongbin and Jung, Dongju and Kim, Kyuri and Chae, Jong Ah and Lim, Jaemin and Park, Jeong Woo and Shin, Doyoon and Char, Kookheon and Jeong, Byeong Guk and Park, Ji-Sang and Hwang, Euyheon and Lee, Doh C. and Park, Young-Shin and Song, Hyung-Jun and Chang, Jun Hyuk and Bae, Wan Ki},
    year = {2023},
    pages = {3779},
}

@article{brokmann_revealing_2009,
    title = {Revealing single emitter spectral dynamics from intensity correlations in an ensemble fluorescence spectrum},
    volume = {17},
    copyright = {© 2009 Optical Society of America},
    number = {6},
    journal = {Optics Express},
    author = {Brokmann, Xavier and Marshall, Lisa and Bawendi, Moungi},
    year = {2009},
    pages = {4509--4517},
}

@article{marshall_extracting_2010,
    title = {Extracting {Spectral} {Dynamics} from {Single} {Chromophores} in {Solution}},
    volume = {105},
    number = {5},
    journal = {Physical Review Letters},
    author = {Marshall, Lisa F. and Cui, Jian and Brokmann, Xavier and Bawendi, Moungi G.},
    year = {2010},
    pages = {053005},
}

@article{cui_direct_2013,
    title = {Direct probe of spectral inhomogeneity reveals synthetic tunability of single-nanocrystal spectral linewidths},
    volume = {5},
    copyright = {2013 Springer Nature Limited},
    number = {7},
    journal = {Nature Chemistry},
    author = {Cui, Jian and Beyler, Andrew P. and Marshall, Lisa F. and Chen, Ou and Harris, Daniel K. and Wanger, Darcy D. and Brokmann, Xavier and Bawendi, Moungi G.},
    year = {2013},
    pages = {602--606},
}

@article{berkinsky_narrow_2023,
    title = {Narrow {Intrinsic} {Line} {Widths} and {Electron}–{Phonon} {Coupling} of {InP} {Colloidal} {Quantum} {Dots}},
    volume = {17},
    number = {4},
    journal = {ACS Nano},
    author = {Berkinsky, David B. and Proppe, Andrew H. and Utzat, Hendrik and Krajewska, Chantalle J. and Sun, Weiwei and Šverko, Tara and Yoo, Jason J. and Chung, Heejae and Won, Yu-Ho and Kim, Taehyung and Jang, Eunjoo and Bawendi, Moungi G.},
    year = {2023},
    pages = {3598--3609},
}

@article{utzat_probing_2017,
    title = {Probing {Linewidths} and {Biexciton} {Quantum} {Yields} of {Single} {Cesium} {Lead} {Halide} {Nanocrystals} in {Solution}},
    volume = {17},
    number = {11},
    journal = {Nano Letters},
    author = {Utzat, Hendrik and Shulenberger, Katherine E. and Achorn, Odin B. and Nasilowski, Michel and Sinclair, Timothy S. and Bawendi, Moungi G.},
    year = {2017},
    pages = {6838--6846},
}

@incollection{webb_fluorescence_2001,
    address = {Berlin, Heidelberg},
    title = {Fluorescence {Correlation} {Spectroscopy}: {Genesis}, {Evolution}, {Maturation} and {Prognosis}},
    shorttitle = {Fluorescence {Correlation} {Spectroscopy}},
    booktitle = {Fluorescence {Correlation} {Spectroscopy}: {Theory} and {Applications}},
    publisher = {Springer},
    author = {Webb, Watt W.},
    editor = {Rigler, Rudolf and Elson, Elliot S.},
    year = {2001},
    pages = {305--330},
}

@article{press_numerical_1993,
    title = {Numerical recipes in {C}: {The} art of scientific computing, second edition},
    volume = {17},
    copyright = {https://www.elsevier.com/tdm/userlicense/1.0/},
    shorttitle = {Numerical recipes in {C}},
    number = {4},
    journal = {Endeavour},
    author = {Press, William H. and Teukolsky, Saul A. and Vetterling, William T. and Flannery, Brian P.},
    year = {1993},
    pages = {201},
}

@article{moerner_dozen_2002,
    title = {A {Dozen} {Years} of {Single}-{Molecule} {Spectroscopy} in {Physics}, {Chemistry}, and {Biophysics}},
    volume = {106},
    number = {5},
    journal = {The Journal of Physical Chemistry B},
    author = {Moerner, W. E.},
    year = {2002},
    pages = {910--927},
}

@article{cai_emission_2024,
    title = {Emission {Mechanism} of {Bright} and {Eco}‐{Friendly} {ZnSeTe} {Quantum} {Dots}},
    volume = {12},
    copyright = {http://onlinelibrary.wiley.com/termsAndConditions\#vor},
    number = {6},
    journal = {Advanced Optical Materials},
    author = {Cai, Wenbing and Ren, Yinjuan and Huang, Zhigao and Sun, Qi and Shen, Hanchen and Wang, Yue},
    year = {2024},
}

@article{jang_synthesis_2019,
    title = {Synthesis of {Alloyed} {ZnSeTe} {Quantum} {Dots} as {Bright}, {Color}-{Pure} {Blue} {Emitters}},
    volume = {11},
    number = {49},
    journal = {ACS Applied Materials \& Interfaces},
    author = {Jang, Eun-Pyo and Han, Chang-Yeol and Lim, Seung-Won and Jo, Jung-Ho and Jo, Dae-Yeon and Lee, Sun-Hyoung and Yoon, Suk-Young and Yang, Heesun},
    year = {2019},
    pages = {46062--46069},
}

@article{lee_crystallographic_2024,
    title = {Crystallographic and {Photophysical} {Analysis} on {Facet}‐{Controlled} {Defect}‐{Free} {Blue}‐{Emitting} {Quantum} {Dots}},
    volume = {36},
    number = {16},
    journal = {Advanced Materials},
    author = {Lee, Yu Jin and Kim, Sungwoo and Lee, Junho and Cho, Eunseog and Won, Yu‐ho and Kim, Taehyung and Kim, Dongho},
    year = {2024},
    pages = {2311719},
}

@article{imran_molecular-additive-assisted_2023,
    title = {Molecular-{Additive}-{Assisted} {Tellurium} {Homogenization} in {ZnSeTe} {Quantum} {Dots}},
    volume = {35},
    copyright = {© 2023 The Authors. Advanced Materials published by Wiley-VCH GmbH},
    number = {45},
    journal = {Advanced Materials},
    author = {Imran, Muhammad and Paritmongkol, Watcharaphol and Mills, Harrison A. and Hassan, Yasser and Zhu, Tong and Wang, Ya-Kun and Liu, Yuan and Wan, Haoyue and Park, So Min and Jung, Euidae and Tam, Jason and Lyu, Quan and Cotella, Giovanni Francesco and Ijaz, Palvasha and Chun, Peter and Hoogland, Sjoerd},
    year = {2023},
    pages = {2303528},
}

@article{han_more_2020,
    title = {More {Than} 9\% {Efficient} {ZnSeTe} {Quantum} {Dot}-{Based} {Blue} {Electroluminescent} {Devices}},
    volume = {5},
    number = {5},
    journal = {ACS Energy Letters},
    author = {Han, Chang-Yeol and Lee, Sun-Hyoung and Song, Seung-Won and Yoon, Suk-Young and Jo, Jung-Ho and Jo, Dae-Yeon and Kim, Hyun-Min and Lee, Bum-Joo and Kim, Hyun-Sik and Yang, Heesun},
    year = {2020},
    pages = {1568--1576},
}

@article{hayee_revealing_2020,
    title = {Revealing multiple classes of stable quantum emitters in hexagonal boron nitride with correlated optical and electron microscopy},
    volume = {19},
    copyright = {2020 The Author(s), under exclusive licence to Springer Nature Limited},
    number = {5},
    journal = {Nature Materials},
    author = {Hayee, Fariah and Yu, Leo and Zhang, Jingyuan Linda and Ciccarino, Christopher J. and Nguyen, Minh and Marshall, Ann F. and Aharonovich, Igor and Vučković, Jelena and Narang, Prineha and Heinz, Tony F. and Dionne, Jennifer A.},
    year = {2020},
    pages = {534--539},
}

@article{park_colloidal_2021,
    title = {Colloidal quantum dot lasers},
    volume = {6},
    copyright = {2021 Springer Nature Limited},
    number = {5},
    journal = {Nature Reviews Materials},
    author = {Park, Young-Shin and Roh, Jeongkyun and Diroll, Benjamin T. and Schaller, Richard D. and Klimov, Victor I.},
    year = {2021},
    pages = {382--401},
}

@article{aharonovich_solid-state_2016,
    title = {Solid-state single-photon emitters},
    volume = {10},
    copyright = {2016 Springer Nature Limited},
    number = {10},
    journal = {Nature Photonics},
    author = {Aharonovich, Igor and Englund, Dirk and Toth, Milos},
    year = {2016},
    pages = {631--641},
}

@article{jang_review_2023,
    title = {Review: {Quantum} {Dot} {Light}-{Emitting} {Diodes}},
    volume = {123},
    shorttitle = {Review},
    number = {8},
    journal = {Chemical Reviews},
    author = {Jang, Eunjoo and Jang, Hyosook},
    year = {2023},
    pages = {4663--4692},
}

@article{angell_unraveling_2024,
    title = {Unraveling sources of emission heterogeneity in {Silicon} {Vacancy} color centers with cryo-cathodoluminescence microscopy},
    volume = {121},
    number = {14},
    journal = {Proceedings of the National Academy of Sciences},
    author = {Angell, Daniel K. and Li, Shuo and Utzat, Hendrik and Thurston, Matti L. S. and Liu, Yin and Dahl, Jeremy and Carlson, Robert and Shen, Zhi-Xun and Melosh, Nicholas and Sinclair, Robert and Dionne, Jennifer A.},
    year = {2024},
    pages = {e2308247121},
}

@article{becker_long_2018,
    title = {Long {Exciton} {Dephasing} {Time} and {Coherent} {Phonon} {Coupling} in {CsPbBr2Cl} {Perovskite} {Nanocrystals}},
    volume = {18},
    number = {12},
    journal = {Nano Letters},
    author = {Becker, Michael A. and Scarpelli, Lorenzo and Nedelcu, Georgian and Rainò, Gabriele and Masia, Francesco and Borri, Paola and Stöferle, Thilo and Kovalenko, Maksym V. and Langbein, Wolfgang and Mahrt, Rainer F.},
    year = {2018},
    pages = {7546--7551},
}

@incollection{empedocles_photoluminescence_1999,
    title = {Photoluminescence from {Single} {Semiconductor} {Nanostructures}},
    copyright = {Copyright © 2000 Wiley-VCH Verlag GmbH},
    booktitle = {Characterization of {Nanophase} {Materials}},
    publisher = {John Wiley \& Sons, Ltd},
    author = {Empedocles, Stephen and Neuhauser, Robert and Shimizu, Kentaro and Bawendi, Moungi},
    year = {1999},
    pages = {261--287},
}

@article{mooney_challenge_2013,
    title = {Challenge to the deep-trap model of the surface in semiconductor nanocrystals},
    volume = {87},
    number = {8},
    journal = {Physical Review B},
    author = {Mooney, Jonathan and Krause, Michael M. and Saari, Jonathan I. and Kambhampati, Patanjali},
    year = {2013},
    pages = {081201},
}

@article{hinterding_single_2021,
    title = {Single {Trap} {States} in {Single} {CdSe} {Nanoplatelets}},
    volume = {15},
    number = {4},
    journal = {ACS Nano},
    author = {Hinterding, Stijn O. M. and Salzmann, Bastiaan B. V. and Vonk, Sander J. W. and Vanmaekelbergh, Daniel and Weckhuysen, Bert M. and Hutter, Eline M. and Rabouw, Freddy T.},
    year = {2021},
    pages = {7216--7225},
}

@article{tan_challenges_nodate,
    title = {Challenges of {II}-{VI} and {III}-{V} {Blue} {Quantum} {Dot} {Light}-{Emitting} {Diodes}},
    volume = {n/a},
    copyright = {© 2025 The Author(s). Advanced Materials published by Wiley-VCH GmbH},
    number = {n/a},
    journal = {Advanced Materials},
    author = {Tan, Shaun and Horowitz, Jonah R. and Tye, Oliver J. and Bawendi, Moungi G.},
    pages = {e12379},
}

@article{yu_comprehensive_2021,
    title = {A {Comprehensive} {Review} of {Fluorescence} {Correlation} {Spectroscopy}},
    volume = {9},
    journal = {Frontiers in Physics},
    author = {Yu, Lan and Lei, Yunze and Ma, Ying and Liu, Min and Zheng, Juanjuan and Dan, Dan and Gao, Peng},
    year = {2021},
}

@article{ahn_self-absorption_2007,
    title = {Self-absorption correction for solid-state photoluminescence quantum yields obtained from integrating sphere measurements},
    volume = {78},
    number = {8},
    journal = {Review of Scientific Instruments},
    author = {Ahn, Tai-Sang and Al-Kaysi, Rabih O. and Müller, Astrid M. and Wentz, Katherine M. and Bardeen, Christopher J.},
    year = {2007},
    pages = {086105},
}

@article{de_mello_improved_1997,
    title = {An improved experimental determination of external photoluminescence quantum efficiency},
    volume = {9},
    copyright = {Copyright © 1997 Verlag GmbH \& Co. KGaA, Weinheim},
    number = {3},
    journal = {Advanced Materials},
    author = {de Mello, John C. and Wittmann, H. Felix and Friend, Richard H.},
    year = {1997},
    pages = {230--232},
}

@article{lelek_single-molecule_2021,
    title = {Single-molecule localization microscopy},
    volume = {1},
    copyright = {2021 Springer Nature Limited},
    number = {1},
    journal = {Nature Reviews Methods Primers},
    author = {Lelek, Mickaël and Gyparaki, Melina T. and Beliu, Gerti and Schueder, Florian and Griffié, Juliette and Manley, Suliana and Jungmann, Ralf and Sauer, Markus and Lakadamyali, Melike and Zimmer, Christophe},
    year = {2021},
    pages = {39},
}

@article{xiao_scanning_2005,
    title = {Scanning {Fluorescence} {Correlation} {Spectroscopy}: {A} {Tool} for {Probing} {Microsecond} {Dynamics} of {Surface}-{Bound} {Fluorescent} {Species}},
    volume = {77},
    shorttitle = {Scanning {Fluorescence} {Correlation} {Spectroscopy}},
    number = {1},
    journal = {Analytical Chemistry},
    author = {Xiao, Ying and Buschmann, Volker and Weston, Kenneth D.},
    year = {2005},
    pages = {36--46},
}

\end{document}